# αβDCA method identifies unspecific binding but specific disruption of the group I intron by the StpA chaperone

Vladimir Reinharz[1,2], Tsvi Tlusty[1*]

[1]Center for Soft and Living Matter, Institute for Basic Science
Ulsan, Republic of Korea

[2]Department of Computer Science, Université du Québec à Montréal,
Montréal, Canada

[*] tsvitlusty@gmail.com

October 10, 2019

Chaperone proteins — the most disordered among all protein groups — help RNAs fold into their functional structure by destabilizing misfolded configurations or stabilizing the functional ones. But disentangling the mechanism underlying RNA chaperoning is challenging, mostly due to inherent disorder of the chaperones and the transient nature of their interactions with RNA. In particular, it is unclear how specific the interactions are and what role is played by amino acid charge and polarity patterns. Here, we address these questions in the RNA chaperone StpA. We adapted direct coupling analysis (DCA) into the αβDCA method that can treat in tandem sequences written in two alphabets, nucleotides and amino acids. With αβDCA, we could analyze StpA-RNA interactions and show consistency with a previously proposed two-pronged mechanism: StpA disrupts *specific* positions in the group I intron while *globally* and loosely binding to the entire structure. Moreover, the interactions are strongly associated with the charge pattern: negatively charged regions in the destabilizing StpA N-terminal affect a few specific positions in the RNA, located in stems and in the pseudoknot. In contrast, positive regions in the C-terminal contain strongly coupled amino acids that promote non-specific or weakly-specific binding to the RNA. The present study opens new avenues to examine the functions of disordered proteins and to design disruptive proteins based on their charge patterns.

**Short title:** Unspecific binding specific disruption of the GII
**keywords:** Group I intron, StpA chaperone, Direct Coupling Analysis, RNA structure, Disordered protein

# 1 Introduction

There is mounting evidence for the existence of Intrinsically Disordered Proteins (IDPs) that lack specific structures (Babu, Kriwacki, and Pappu 2012). These proteins do not fold into a well-defined conformation (Wright and Dyson 2015), although some may acquire a specific structure given the right context. IDPs are at the core of key biological assemblies and processes, such as membrane-less organelles (Nott et al. 2015), cell signaling (Wright and Dyson 2015), and cell division (Buske and Levin 2013). Disordered regions may exert entropic forces on the proteins they bind and thereby shift the ensemble of protein structures towards one with higher binding affinity (Keul et al. 2018). While our repertoire of IDPs is steadily growing (Varadi and Tompa 2015; Schad et al. 2017; Piovesan et al. 2017), the function of most is yet to be discovered (Van Der Lee et al. 2014; Papaleo et al. 2016). Nevertheless, analysis suggests that a crucial determinant of the global shape and function of IDPs is their charge pattern (Das, Ruff, and Pappu 2015).

A prominent class of IDPs is that of chaperones whose fraction of disordered residues, 54% on average, is the highest among all functional classes of proteins (Tompa and Csermely 2004). A particularly important subclass is those that chaperone RNA folding: To perform their functions, non-coding RNAs rely on well-conserved structures, which have been used for sequence alignment and putative RNA prediction (Nawrocki and Eddy 2013). While some non-coding RNAs are able to attain those structures by themselves, chaperone proteins are essential in stabilizing correct conformations or in destabilizing, and thus rescuing, misfolded RNAs (Bhaskaran and Russell 2007; Woodson 2010; Papasaikas and Valcárcel 2016).

A prime example of chaperone-dependent RNA is the Group I Intron (GII), which has an elaborate functional structure (Michel and Westhof 1990). Two chaperones take part in the folding of this RNA. One is the Cyt-18 protein that stabilizes the active structure (Mohr et al. 1992; Guo and Lambowitz 1992). The second chaperone is the `StpA` protein, which is known to destabilize misfolded GII structure (Waldsich, Grossberger, and Schroeder 2002; Mayer et al. 2007). The structures of Cyt-18 and its complex with the GII are well-determined (Paukstelis et al. 2008). In contrast, most of the `StpA` protein, 73% of the residues, is known to be disordered. `StpA` consists of two domains, the N-terminal and C-terminal. Excising the C-terminal from the sequence increases the efficacy of the chaperone, while mutations in the C-terminal hinder its binding capacity (Mayer et al. 2007). An entropy transfer model has been proposed, where rapid and transient binding disturbs the structure, thus allowing it to refold (Tompa and Csermely 2004). But many questions regarding the specifics of the destabilization function remain open. An inherent obstacle in understanding the mechanisms of disordered proteins, such as `StpA`, is the lack of functional structure. The `StpA`-GII problem is even more challenging since the other partner in the interaction, the GII RNA, is misfolded and therefore lacks a specific structure as well.

To overcome the lack of structures, one may leverage the accelerated growth in the number of known sequences and use them for multiple sequence alignments (MSA). As of August 2018, `GenBank` (Sayers et al. 2019) had sequences totaling over nucleotides from species, an increase of 40% from the previous year. Techniques such as Direct Coupling Analysis (DCA) extract from the MSAs amino acid contacts and 3D structures (Burger and Van Nimwegen 2010; Ovchinnikov et al. 2015; Marks et al. 2011), protein–protein interaction sites (Morcos et al. 2011; Ovchinnikov et al. 2014), RNA ligand binding pockets (Reinharz et al. 2016), RNA tertiary contacts (De Leonardis et al. 2015), and RNA–protein interaction sites (Weinreb et al. 2016). These studies have also demonstrated that many IDPs have strong correlations, hinting at context dependant structures (Toth-Petroczy et al. 2016), though in the last study `StpA` did not exhibit any particular structure. So far, however, IDP–RNA interactions — which are essential in many molecular systems, in particular chaperones — have not been examined, perhaps due to the difficulty of analyzing the interaction of two objects that lack defined structures and whose sequences are written in different alphabets.

All this motivates the present study in which we adapt the DCA method to concurrently process proteins and RNAs, which not only differ in the size of their alphabets but, on top of that, have high variability in sequence conservation. The adaptive method, termed αβDCA, produces the first analysis of the interaction of a disordered protein, `StpA`, with a non-coding RNA, the group I intron. Our method identifies 90 strongly coupled pairs between `StpA` and GII. The inferred locations of those pairs are consistent with the results of (Mayer et al. 2007).

We find that the charge pattern is strongly associated with the type of interactions: The N-terminal of `StpA`, which is known to destabilize the RNA, exhibits a few specific interactions among negatively charged regions of the protein and regions of the GII, which are critically misfolded in the structures ensemble or impede functional loops from forming. In the C-terminal, strongly coupled amino acids are mostly in positively charged regions, and their interaction of these amino acids with the RNA is weakly specific and almost uniformly distributed over the entire GII sequence. Moreover, while both terminals are of roughly the same length, only 21% of the top DCA scores are in the N-terminal. These findings propose a charge-dependent two-pronged mechanism of unspecific binding but specific disruption by chaperone IDPs.

## 2  Results

We extended the classic mean-field approximation DCA (mfDCA) method for treating paired sequences that are written in different alphabets and have different levels of sequence conservation (for details see Sec. 4). First, we tested this simple DCA variant – which we call αβDCA (for treating varying alphabets) – against two other DCA implementations: Gremlin, an implementation of Markov random-field DCA (Ovchinnikov et al. 2014), and EVcouplings (Hopf et al. 2018), an implementation of pseudo-likelihood DCA (plmDCA). For the benchmark of the 5S–RL18 ribosomal complex, the adaptive αβDCA method predicts more contacts in its top scores (see Sec. 4.5). Additionally, we observe that the mfDCA method outperforms Gremlin in the GII alignment, most probably owing to the correct pseudo-count for a 5-letter alphabet, rather than that of the 21-letter alphabet of proteins used in Gremlin. In the following, we apply the αβDCA method to analyze the StpA–GII alignment (code and alignment are available at: https://gitlab.info.uqam.ca/cbe/abDCA).

## 2.1 αβDCA exhibits significant scores for strongly-coupled StpA–RNA contacts

The DCA method identifies strong couplings, indicating significant physical interactions. These significant scores emerge as outliers departing from the bulk distribution of the DCA scores. Sequence conservation is a critical factor, as too high conservation level prohibits co-evolution analysis. Fig. 1a shows the secondary structure of the GII RNA together with its long range interactions and sequence conservation values (the overall maximal conservation is shown in Fig. 3). To test whether the alignment contains more information than an ensemble of random sequences, we compare the distribution of αβDCA scores from the StpA-GII alignment with those obtained from the same alignment but with randomly shuffled sequences. The scores of the original alignment spread over a much wider range then that of the shuffled alignment, thus confirming that the DCA analysis extracts information from the alignment (Fig. 5).

The StpA–GII amino acid-nucleotide pairs with the strongest DCA couplings are shown in Fig. 1b. The distribution of scores is assumed to be normal, with its average and standard deviation computed from the empirical data. Scores that are 4 standard deviations ($4\sigma$) above the average are deemed significant. The αβDCA identifies 90 significant pairs, 15% less than those extracted by the standard DCA, which disregards the difference in alphabet and sequence similarity between the RNAs and the proteins. As shown below, in agreement with previous studies (Ovchinnikov et al. 2015; Toth-Petroczy et al. 2016), the number of false positives increases with the number of selected pairs. One can therefore expect that the analysis that shows fewer significant scores will yield fewer errors.

## 2.2 Inferred protein-RNA interactions are selective in the N-terminal and global in the C-terminal of StpA

The N- and C-terminals of StpA are known to interact differently with the RNA (Waldsich et al. 2002). This motivates us to characterize the number and distribution of high αβDCA scores, which indicate strong physical couplings, in each of these two regions. Since RNA structures fluctuate within a dynamic ensemble (McCaskill 1990), we examine the interactions in light of the two main structure ensembles and the functional structure, and in particular link the distribution of strong couplings along the RNA.

To this end, from the RNA sequence, `RNAstructure` (Reuter and Mathews 2010) computes, in the McCaskill thermodynamic framework (McCaskill 1990), the probability of each possible base pair. Those pairing probabilities can be divided into two main structural ensembles to ease the visualization (Aalberts and Jannen 2013). We plot in Fig. 2 the net charge distribution along the `StpA` protein (averaged over a window of 5 amino acids), above the two main clusters of the GII RNA structure ensemble as predicted by `RNAstructure` (Reuter and Mathews 2010). The arcs in the upper part depict bonds in the main cluster, whose probability is 68.2%, and the arcs in the lower part show bonds in the second main cluster, of probability 31.8% (Aalberts and Jannen 2013). The red discs represent stems in the functional structure that are absent from both ensembles, in particular the pseudoknot, as annotated by (Waldsich et al. 2002) . Note that while pseudoknots cannot be predicted with `RNAstructure`, they could not be inferred even with `RNAPKplex`, which was designed for this purpose (Lorenz et al. 2011).

The significant scores between the RNA and the protein are denoted by lines. There are 90 significant scores ($\geq 4\sigma$) between the protein and the RNA: 19 in negative regions of the N-terminal (dark lines), 69 in mostly positive regions of the C-terminal (light gray lines), and 2 in the linker between the N and C-terminals (dashed lines).

More than 61% of C-terminal significant amino acids have many globally distributed partners, on average 4.3 nucleotides. In contrast, the N-amino acids shows a more selective evolutionary signature with 67% of them exhibiting significant co-variation with *only* one nucleotide.

## 2.3  High scores correspond to close nucleotides in the 3D structure of GII

To check whether the RNA alignment is informative by itself, we examine the DCA scores among all pairs of RNA positions. To validate the quality of the RNA alignment, we compared the physical contacts predicted by DCA to the 3D structure of the *td* GII RNA (available at http://www-ibmc.u-strasbg.fr/spip-arn/spip.php?rubrique136). We computed DCA scores using two methods, the mean-field approximation (mfDCA) and Gremlin (Ovchinnikov et al. 2014). We note that αβDCA is identical to mfDCA when treating a single alphabet. We consider as a good prediction a pair of nucleotides closer than 8Å in the 3D structure. Fig. 4 shows the number of these true positives (distance < 8Å) for the hundred top scores. While the first 40 top scores are well predicted by both methods, the Gremlin method is outperformed by mfDCA in the next 60 scores.

# 3 Discussion

The `StpA` protein destabilizes the misfolded GII RNA, allowing it to achieve its functional structure. Experiments have shown that the binding is transient and weak, with little specificity (Waldsich et al. 2002; Doetsch et al. 2011). Mutation studies provide evidence for GII-`StpA` interactions: Mutations in the `StpA` C-terminal reduce the binding affinity between `StpA` and the group I intron, while complete deletion of the C-terminal increases the efficiency of `StpA` as a chaperone (Waldsich, Grossberger, and Schroeder 2002). A C-terminal mutation, glycine 126 changed to valine, weakens the binding and increases the efficiency of `StpA` (Mayer et al. 2007). In the following, we further expand the understanding of the GII-`StpA` mechanism, based on the αβDCA results. We show that **the αβDCA results** are consistent with previous experimental studies. **Moreover, they put forward** a detailed picture of coupled amino acids and nucleotides responsible for both binding and destabilizing interactions.

## 3.1 Binding is mediated by positively charged regions of StpA

Binding of `StpA` to GII is driven by electrostatic forces mediated by positively charged amino acids (Mayer et al. 2007). This is confirmed by the αβDCA showing that the vast majority of high scores in the C-terminal are in positively charged regions (Fig. 2). Out of the 69 pairs, 44 (64%) are in positively charged regions, 18 (26%) in neutral regions and 7 (10%) in negatively charged regions. This also implies that most of the binding energy comes from amino acids in the C-terminal. It was conjectured that binding is only weakly specific and prefers unstructured RNAs (Mayer et al. 2007). Our analysis is consistent with this conjecture, showing a spread of top αβDCA scores all over the RNA. Fig. 7 shows the cumulative number of top scores of with the C-terminal along the RNA, demonstrating the roughly uniform spread (with gaps excluded), with notable enrichment before position 200.

In a fine-grained examination, one notices several interactions of special interest. The glycine at position 126 of the protein, which is known to strongly reduce binding affinity when mutated, takes part in three different pairs. Position 113 of the protein — which participates in 14 different pairs, more than any other amino acid — is strongly coupled to positions 125 and 162 in the RNA, which themselves are also coupled with glycine 126. Position 125 of the RNA resides in the 5′-end of the pseudoknot, and position 162 in the 3′-end of the P3 stem. The two RNA regions with the strongest coupling to the C-terminal are both ends of the pseudoknot, which are involved in erroneous base pairs in the two dominant structures. This may explain why the isolated C-terminal is a much inferior chaperone than the whole protein. Our analysis is also consistent with the theory that while important misfolded regions are disrupted, strong electrostatic binding slows the release of `StpA`, thereby impeding the correct folding of the RNA.

## 3.2 Destabilization is mediated by negatively charged regions targeting specific RNA positions

Removing the linker and C-terminal increases by 50% the efficiency of `StpA`, implying that the N-terminal drives the destabilization (Mayer et al. 2007). While the N-terminal composes 48.5% of `StpA`, it contains only 21% of the strong couplings, 19 out of 90. The black lines in Fig. 2 show the coupled pairs between `StpA` and GII. Out of 19 significant pairs, 1 is in a positively charged regions, 5 (26%) in a neutral region, and the remaining 13 (68%) in negatively charged regions.

Of those 19 pairs, 7 are coupled with regions that determine the functional RNA conformation in both structure ensemble, in particular the one position paired with the positively charged amino acid at position 39 in the N-terminal. The other 12 pairs correlate with four different regions that are expected to be destabilized as the probable structure conflicts with the functional one.

In both ensembles, the functional short stems at the beginning and end of the GII sequence are blocked by a stem linking those two parts together. We find three interactions that target this region: First, the 3′-end of the pseudoknot, in both ensembles, is blocked by misfolded stems which are strongly correlated with a position of the N-terminal. Second, following the 5′-end of the P3 stem, the region between positions 60 and 85 of the RNA has the right conformation in the less probable ensemble and is targeted by three couples. Finally, 3′-end of the P3 stem, present only in the least probable ensemble, is involved in one coupling. The last 3 coupled positions are in a hairpin stem preceding the P3 3′-end. This stem is missing functional base pairs in both ensembles, two of the coupled pairs are in positions lacking a base pair, the third in the unpaired region of the hairpin.

Without `StpA`, around 55% of the RNA is able to fold into its functional self-splicing form, and this folding fraction rises to roughly 80% in the presence of the chaperone (Mayer et al. 2007). The strong correlations we observe manifest an interplay between the two main structure ensembles of the RNA, with the less probable one presenting most of the correct base pairs. Regions that contain functional stems in the least probable ensemble are all targeted by couplings with the destabilizing N-terminal. In both ensembles, the functional but energetically unfavorable pseudoknot has stems in its 3′-end impeding its formation. Our analysis proposes that the stems are also destabilized by the N-terminal.

## 3.3 Conclusion

DCA methods have been applied to infer protein structure, and protein–protein or protein–RNA interactions (Weinreb et al. 2016). DCA demonstrated high correlations among amino acids in IDPs, suggesting that many IDPs do exhibit structure in a particular context (Toth-Petroczy et al. 2016). In the present study, we expanded DCA to account for the different alphabets and different levels of sequence diversity in the concatenated sequences of protein and RNA used for the alignment. We used this adapted αβDCA method to infer the strong couplings between a non-coding RNA, GII, and its disordered protein chaperone, `StpA`. Understanding the `StpA`-GII is particularly challenging, since on top of the inherent disorder of the protein, the misfolded RNA also lacks a well-defined structure.

The present αβDCA method produces 15% less significant contacts than the traditional mfDCA. In cases where the structure is unknown, a rather arbitrary significance threshold must be chosen. Having less scores departing from the distribution indicates better discrimination of important co-evolving pairs. Our findings are consistent with experiments and a proposed mechanism where the binding, mediated by electrostatic forces of positively charged amino acids, is *non-specific* or only weakly specific. These strong couplings, observed in the positively charged regions in the C-terminal of `StpA`, are paired with evenly distributed nucleotides along the RNA sequence. In contrast, the αβDCA suggests that the structural disruption driven by the N-terminal is mediated by negatively amino acids that target *specific* regions of the RNA sequence. In particular, regions in the two main structure ensembles of the RNA impeding the formation of the first and last stem, as the pseudoknot, are strongly coupled with the N-terminal. Stems in the more probable structure ensemble — which are conflicting with the functional stems present in the lower probability ensemble — are also targeted.

The present study is the first direct coupling analysis of the coupling between a disordered chaperone and its RNA target. Charge patterns have been known to be crucial for the global structure of disordered proteins, and here we shed some light on how they can affect destabilization mechanisms involved in RNA chaperoning. The analysis suggests several concrete experimental tests, for example mutations at positions 99 and 113 in the C-terminal are expected to significantly decrease binding affinity. The αβDCA variant used in the study is simple and general enough to be easily applied for investigating other IDP-RNA mechanisms. An interesting application of the present analysis is the identification of chaperone IDPs from their charge pattern. Those patterns could also be used to design novel destabilizing proteins.

# 4  Method

We first present a modified DCA algorithm, termed αβDCA, adapted for treating paired sequences that are written in different alphabets and have different sequence conservation levels. The different nature of the paired sequences influences the normalization factors that are crucial to predict the disentangled covariations. To illustrate the method, we show how the data for the `StpA` protein and the group I intron RNA was gathered, and how the alignment was built. The code is freely available at: https://gitlab.info.uqam.ca/cbe/abDCA.

## 4.1  αβDCA: direct coupling analysis for varying alphabets and sequence conservation

Direct coupling analysis (DCA) has proved extremely useful for disentangling covariations between non-interacting residues in multiple sequence alignment (MSA) (Weigt et al. 2009; Morcos et al. 2011). It aims to find the Potts model that maximizes the entropy, in order to infer the most likely probability having the given dinucleotide marginals without any additional constraints (Weigt et al. 2009). The original method was constructed to treat alignments of sequences written in the same alphabet, namely the protein amino acids written in the language of the genetic code. We modify this method to treat in tandem two alphabets, of sizes $r$ and $s$. Given a sequence of $n$ characters, we assume that the first $\zeta$ elements are from the alphabet of size $r$, and the last $n - \zeta$ from the alphabet of size $s$. In this study, the first alphabet is of the protein amino acids and a gap, hence $r = 21$, and the second is of the RNA nucleotides and a gap, i.e. $s = 5$.

The MSA of $M$ sequences of length $n$ is recorded as its sequence of columns $\{C_1^p, \dots, C_n^p\}$ where $p \in [1, \dots, M]$ are the $M$ sequences and $1, \dots, n$ are the columns. Since the proteins and RNAs have different sequence similarity and alphabets, we define two values for calibrating the pseudo-count:

$$m_p^{\text{prot}} = \sum_{q=1}^{M} \left[ 1 \text{ if similarity}(C_{1,\dots,\zeta}^p, C_{1,\dots,\zeta}^q) > 80\% \right],$$

$$m_p^{\text{rna}} = \sum_{q=1}^{M} \left[ 1 \text{ if similarity}(C_{\zeta+1,\dots,n}^p, C_{\zeta+1,\dots,n}^q) > 80\% \right],$$

where similarity$(C_{a,\dots,b}^p, C_{a,\dots,b}^q) > 80\%$ is true if sequences $C^p$ and $C^q$ are identical in over 80% of the positions between $a$ and $b$. We note that the values of $m_p^{\text{prot}}$ and $m_p^{\text{rna}}$ are at least 1 since each sequence is identical to itself. We additionally define $M_{\text{eff}}^{\text{prot}} = \sum_{p=1}^{M} 1/m_p^{\text{prot}}$ and

$M_{\text{eff}}^{\text{rna}} = \sum_{p=1}^{M} 1/m_p^{\text{rna}}$. The parameter $\lambda$ is a pseudocount set to the appropriate value of $M_{\text{eff}}^{\text{prot}}$ or $M_{\text{eff}}^{\text{rna}}$, as in previous studies.

The frequencies of each letter in each column, and of each pair of letters for each pair of positions, need to be re-weighted as following. We define the frequency count of a letter $\alpha$ at column $i$, given the indicator function $\mathbb{1}$, as:

$$f_i(\alpha) = \begin{cases} \dfrac{1}{M_{eff}^{prot} + \lambda} \left( \dfrac{\lambda}{r} + \sum_{p=1}^{M} \dfrac{1}{m_p^{prot}} \mathbb{1}_{\alpha, c_i^p} \right) : i \leq \zeta \\ \dfrac{1}{M_{eff}^{rna} + \lambda} \left( \dfrac{\lambda}{s} + \sum_{p=1}^{M} \dfrac{1}{m_p^{rna}} \mathbb{1}_{\alpha, c_i^p} \right) : i > \zeta \end{cases}$$

Similarly, the frequency count of a pair of letters $(\alpha, \beta)$ at positions $(i, j)$ is defined as:

$$f_{i,j}(\alpha, \beta) = \begin{cases} \dfrac{1}{M_{eff}^{prot} + \lambda} \left( \dfrac{\lambda}{r^2} + \sum_{p=1}^{M} \dfrac{1}{m_p^{prot}} \mathbb{1}_{\alpha, c_i^p} \mathbb{1}_{\beta, c_j^p} \right) : i < j \leq \zeta \\ \dfrac{1}{\frac{1}{2}(M_{eff}^{prot} + M_{eff}^{rna}) + \lambda} \left( \dfrac{\lambda}{rs} + \sum_{p=1}^{M} \dfrac{1}{\frac{1}{2}(m_p^{prot} + m_p^{rna})} \mathbb{1}_{\alpha, c_i^p} \mathbb{1}_{\beta, c_j^p} \right) : i \leq \zeta < j \\ \dfrac{1}{M_{eff}^{rna} + \lambda} \left( \dfrac{\lambda}{s^2} + \sum_{p=1}^{M} \dfrac{1}{m_p^{rna}} \mathbb{1}_{\alpha, c_i^p} \mathbb{1}_{\beta, c_j^p} \right) : \zeta < i < j \end{cases}$$

The rest of the equations follow closely the formulation in (Morcos et al. 2011). The coupling value $e_{ij}(\alpha, \beta)$, between two letters $(\alpha, \beta)$ at positions $(i, j)$, is calculated through the set of $n(n-1)/2$ matrices $\partial$, the connected correlation matrix. For each pair of positions $i, j$, one defines a matrix $\partial_{ij}$, whose dimension is: $(r-1)^2$ if $i < j \leq \zeta$, $(r-1)(s-1)$ if $i \leq \zeta < j$, and $(s-1)^2$ if $\zeta < i < j$. For all $i \in [1, \ldots, n], j \in [1, \ldots, n]$ the entries of $\partial_{ij}$ are:

$$\partial_{ij}(\alpha, \beta) = f_{ij}(\alpha, \beta) - f_i(\alpha) f_j(\beta)$$

where $\alpha$ and $\beta$ take all possible $r - 1$ or $s - 1$ values, depending on the index $i$ and $j$. Finally, the coupling between positions $i, j$ is obtained by inverting $\partial$

$$e_{ij} = -\left(\partial_{ij}^{-1}\right)$$

where that block matrix is extended with 0s so that the dimension of $e_{ij}$ is $r^2$ if $i < j \leq \zeta$, $rs$ if $i \leq \zeta < j$, and $s^2$ if $\zeta < i < j$. The inverse of the connected correlation matrix returns the negative coupling term, we correct it by taking minus its value (Morcos et al. 2011).

We can now define a pseudo-probability, $P_{ij}(\alpha,\beta)$, of observing $(\alpha,\beta)$ at positions $(i,j)$, given auxiliary residue fields $\tilde{h}$ for each position:

$$P_{ij}(\alpha,\beta) = \frac{1}{\mathcal{Z}}\exp\bigl[e_{ij}(\alpha,\beta) + \tilde{h}_i(\alpha) + \tilde{h}_j(\beta)\bigr],$$

where $\mathcal{Z}$ is the normalization factor. The values of the fields $\tilde{h}$ are determined by the observed single residue count, and must satisfy the system of equations:

$$f_i(\alpha) = \sum_\gamma P_{ij}(\alpha,\gamma),\ f_j(\beta) = \sum_\gamma P_{i,j}(\gamma,\beta),$$

noting we must assume that if $i \leq \zeta: \tilde{h}_i(r) = 0$ (resp. if $\zeta < i: \tilde{h}_i(s) = 0$).

At this point, we can compute the directed information between two positions, $D_{ij}$, as:

$$D_{ij} = \sum_{\alpha,\beta} P_{ij}(\alpha,\beta)\ln\frac{P_{ij}(\alpha,\beta)}{f_i(\alpha)f_j(\beta)}.$$

Finally, the distortion of the scores due to the undersampling effect is corrected using an average product correction (APC) method (Dunn, Wahl, and Gloor 2007).

## 4.2 StpA homologues

The `StpA` protein from *Escherichia coli* (strain K12) sequence is:
MSVMLQSLNNIRTLRAMAREFSIDVLEEMLEKFRVVTKERREEEEQQQRELAERQEKIST
WLELMKADGINPEELLGNSSAAAPRAGKKRQPRPAKYKFTDVNGETKTWTGQGRTPKPIA
QALAEGKSLDDFLI.

The distribution of charges along the sequence is a known indicator of the global conformation of disordered proteins (Holehouse et al. 2017). The Das–Pappu phase diagram shows that the `StpA` protein belongs to the ensemble of "Janus sequences". Those are collapsed or expanded depending of context, and most functional disordered proteins belong to that group. This region of Janus sequences contains 40% of known disordered proteins (Das et al. 2015), whereas another 25% reside in the strong polyampholyte region, and 30% are classified as weak polyampholyte.

The `jackhmmer` method (Potter et al. 2018) was run iteratively 13 times, until the number of sequences added to the matches was less than 1% of the already identified ones. We identified 21593 matches, 5749 of them unique. `jackhmmer` provides a sequence alignment of all the hits, which belong to 7539 different taxa. Every sequence in GenBank (Sayers et al. 2019) associated with those taxa was downloaded, a total of 633GB of data.

## 4.3 Group I intron

The *td* group I intron (GII) sequence from *phage T4 thymidylate-synthase* is:

gguUAAUUGAGGCCUGAGUAUAAGGUGACUUAUACUUGUAAUCUAUCUAAACGGGGAACC
UCUCUAGUAGACAAUCCCGUGCUAAAUUGUAGGACUGCCCGGGUUCUACAUAAAUGCCUA
ACGACUAUCCCUUUGGGGAGUAGGGUCAAGUGACUCGAAACGAUAGACAACUUGCUUUAA
CAAGUUGGAGAUAUAGUCUGCUCUGCAUGGUGACAUGCAGCUGGAUAUAAUUCCGGGGUA
AGAUUAACGACCUUAUCUGAACAUAAUGcuac

and its functional secondary structure, from (Waldsich et al. 2002), is

((((......))))((((((((((....))))))))))...(((((...(((((....(((.....)))...))))))((.....(((((((((.....))))))))).....))...[.[[[[.(((....)))(.(((((((....)))))))..))))))(((((......))))))))......]]]]]...(((((((....))))))).((((......))))(((((((..........))))))))..............

where the pseudoknot is indicated with square brackets, '[' and ']'.

The GII has 14 different subgroups, which have been cataloged in the GISSD database (Zhou et al. 2008). Identification and alignment of GII sequences are highly dependent of the subgroup they belong to (Nawrocki et al. 2018). Therefore, for each subgroup, we generated a covariance model using Infernal (Nawrocki and Eddy 2013). The IA2 subgroup is the most compatible with GII. With GII, Infernal reports an E-value of $1.7 \times 10^{-36}$ and 63% of the base pairs are well predicted. In particular, the complete P3 stem (brown in Fig. 2) is perfectly aligned with the consensus structure. We note that while the sequence has 273 nucleotides, only the first 248 were matched. The rest of the analysis is performed on those 248 nucleotides.

A search of matches to the IA2 subgroup was then computed with the cmsearch routine of Infernal, on all sequences from the 7359 taxa gathered previously. A total of 7542 sequences were identified as significant — e-value < 0.01 — with default parameters, 471 of them unique. The cmsearch tool returns an alignment of those sequences.

## 4.4 Protein–RNA alignment

Duplicate proteins and RNAs were removed from each taxon. Every possible protein–RNA pair inside a taxon was concatenated together. This yielded a total of 13 230 couples, of which 10 013 unique.

Only columns where `StpA` and the GII have less than 50% of gaps were kept. In total, 39 positions of the proteins were removed, the N-terminal's first 30 positions, 6 in the C-terminal and 2 in the linker. In the RNA, 64 positions were removed. The resulting protein alignment is composed of 95 columns and the RNA alignment of 184.

### 4.5  5S RNA–RL18 protein interactions

We compare four DCA methods for the benchmark of inferring the interactions between the 5S RNA and the RL18 protein. The four methods are: (i) standard mfDCA, where the pseudocount is kept at 21 for every position in our alignment, (ii) Our αβDCA implementation of mfDCA with adaptive pseudocount, (iii) The implementation EVcouplings (Hopf et al. 2018) of pseudo-likelihood DCA (plmDCA), and (vi) The Markov-random field DCA as implemented in Gremlin (Ovchinnikov et al. 2014).

We used the protein alignment of RL18 provided in (Weinreb et al. 2016). The RNA sequences where recovered from the Rfam family `RF00001` (Kalvari et al. 2017). We followed the protocol of Sec. 4.4. Due to the large amount of sequences, we selected randomly one pair of protein–RNA per taxonomic family, as in (Weinreb et al. 2016). The alignment before removing columns with over 50% of gaps is available at: https://gitlab.info.uqam.ca/cbe/abDCA.

We computed amino acid–nucleotide distances in the `4V4Q` protein structure (Schuwirth et al. 2005). Pairs with distance shorter than 10Å are considered to be in contact.

We show in Fig. 6 the results of the first top 100 scores for each method. Only mfDCA and αβDCA (mfDCA adaptive) have their highest scores correctly predicting a contact. While mfDCA's fourth hit is correct but not the one in αβDCA method, the opposite occurs at their sixth top score. Both methods outperform Gremlin and EVcouplings on the top 20 scores. While mfDCA and αβDCA true positives steadily declines as more top scores are taken into account, Gremlin sees an increase to up to 50% at its 30$^{th}$ score. All methods then converge to roughly 22% true positive when the first 100 scores are taken into account.

The overlap of scores over 4$\sigma$ from each bulk distribution is shown in Fig. 8 (Heberle et al. 2015). While 95% of those overlap between mfDCA and αβDCA, they are almost completely exclusive from Gremlin and EVcouplings top results. None of the top pairs is identified by all of the four methods and only 2 are shared by mfDCA, αβDCA and Gremlin. This is the only overlap between any three methods.

## 5 Acknowledgments

We thank Thomas Hopf and Debora S. Marks for help with the EVcoupling suite, Sergey Ovchinnikov for providing Gremlin, Eric Westhof for providing the structure of the td-Intron, and Eric Nawrocki for help with infernal and understanding how to generate the subgroup GII alignments. This work was supported by the taxpayers of South Korea through the Institute for Basic Science, project code IBS-R020. VR was also supported by the Natural Sciences and Engineering Research Council of Canada (NSERC) [RGPIN-2020-05795].

## 6 Captions

**Caption Fig1.** **a.** GII secondary structure and its sequence conservation. The last 25 positions have no conservation levels since they are excluded from the alignment (see Sec. 4.3). **b.** Positions of significant DCA scores ($\geq 4\sigma$ above average) between the `StpA` protein (vertical axis) and the GII RNA (horizontal axis). The RNA axis is labeled with the secondary structure in parentheses notation. The blue region is the N-terminal of `StpA` and the orange region its C-terminal. The pale grey regions are the pseudoknot (PK) of GII. The last 25 positions of GII are omitted (See Sec. 4.3).

**Caption Fig2.** Significant scores between the protein and the RNA. Top: the protein charge distribution. Bottom: the RNA sequence with the two main structure clusters. Arcs represent base pairs: yellow only in the main, most probable cluster, blue only in the secondary, least probable cluster, and black in the functional structure. Red discs are base pairs in the functional structure absent from both clusters. Brown arcs are the P3 stem. Positions highlighted in green in the protein and RNA had more than 50% of gaps in the alignment and were therefore omitted from the analysis. Significant DCA scores are denoted by lines: dark between the N-terminal and the RNA, light gray lines between the C-terminal and the RNA.

**Caption Fig3. Global conservation** The most conserved nucleotide for each position, with its percentage of conservation. Each nucleotide shown is the most frequent one. If only `A` or `G` are present in that position, an `R` is shown for purine. If only `C` or `U` are present in that position, a `Y` is shown for pyrimidine.

**Caption Fig4.** Evaluating the predictive power of the group I intron RNA sequence alignment. Fraction of nucleotide pairs closer than 8Å for the top DCA values using mfDCA and Gremlin.

**Caption Fig5.** The distribution of APC values obtained by our method on the `StpA`-GII alignment compared to the same values after shuffling the sequences. There are 100 blue and 100 orange bins. The orange bins are therefore narrower.

**Caption Fig6.** Comparing four DCA methods for the benchmark of inferring the 5S–RL18 complex from PDB `4V4Q`. The graphs show fraction of pairs with a distance below 10Å for the top 100 DCA values for each method. The circles indicate the last score over $4\sigma$ from the bulk distribution.

**Caption Fig7.** Cumulative distribution of top DCA scores of amino acids in the C-terminal of `StpA` coupled with nuecletides along the RNA. Positions with over 50% of gaps omitted from the analysis are in grey. The black curve is the cumulative uniform distribution with the same gaps.

**Caption Fig8.** Intersection of the scores over $4\sigma$ for each of the four DCA methods on the 5s–RL18 complex.

# 7 References

Aalberts, Daniel P, and William K Jannen. 2013. "Visualizing RNA Base-Pairing Probabilities with RNAbow Diagrams." *RNA* 19 (4): 475–78.

Babu, M Madan, Richard W Kriwacki, and Rohit V Pappu. 2012. "Versatility from Protein Disorder." *Science* 337 (6101): 1460–1.

Bhaskaran, Hari, and Rick Russell. 2007. "Kinetic Redistribution of Native and MisfoldedRNAs by a DEAD-Box Chaperone." *Nature* 449 (7165): 1014.

Burger, Lukas, and Erik Van Nimwegen. 2010. "Disentangling Direct from Indirect Co-Evolution of Residues in Protein Alignments." *PLoS Computational Biology* 6 (1): e1000633.

Buske, PJ, and Petra Anne Levin. 2013. "A Flexible C-terminal Linker Is Required for Proper FtsZ Assembly in Vitro and Cytokinetic Ring Formation in Vivo." *Molecular Microbiology* 89 (2): 249–63.

Das, Rahul K, Kiersten M Ruff, and Rohit V Pappu. 2015. "Relating Sequence Encoded Information to Form and Function of Intrinsically Disordered Proteins." *Current Opinion in Structural Biology* 32: 102–12.

De Leonardis, Eleonora, Benjamin Lutz, Sebastian Ratz, Simona Cocco, Rémi Monasson, Alexander Schug, and Martin Weigt. 2015. "Direct-Coupling Analysis of Nucleotide

Coevolution Facilitates RNA Secondary and Tertiary Structure Prediction." *Nucleic Acids Research* 43 (21): 10444–55.

Doetsch, Martina, Renée Schroeder, and Boris Fürtig. 2011. "Transient RNA–Protein Interactions in RNA Folding." *The FEBS Journal* 278 (10): 1634–42.

Dunn, Stanley D, Lindi M Wahl, and Gregory B Gloor. 2007. "Mutual Information Without the Influence of Phylogeny or Entropy Dramatically Improves Residue Contact Prediction." *Bioinformatics* 24 (3): 333–40.

Guo, QINGBIN, and Alan M Lambowitz. 1992. "A Tyrosyl-TRNA Synthetase Binds Specifically to the Group I Intron Catalytic Core." *Genes & Development* 6 (8): 1357–72.

Heberle, Henry, Gabriela Vaz Meirelles, Felipe R da Silva, Guilherme P Telles, and Rosane Minghim. 2015. "InteractiVenn: A Web-Based Tool for the Analysis of Sets Through Venn Diagrams." *BMC Bioinformatics* 16 (1): 169.

Holehouse, Alex S, Rahul K Das, James N Ahad, Mary OG Richardson, and Rohit V Pappu. 2017. "CIDER: Resources to Analyze Sequence-Ensemble Relationships of Intrinsically Disordered Proteins." *Biophysical Journal* 112 (1): 16–21.

Hopf, Thomas A, Anna G Green, Benjamin Schubert, Sophia Mersmann, Charlotta PI Schärfe, John B Ingraham, Agnes Toth-Petroczy, et al. 2018. "The EVcouplings Python Framework for Coevolutionary Sequence Analysis." *Bioinformatics* 35 (9): 1582–4.

Kalvari, Ioanna, Joanna Argasinska, Natalia Quinones-Olvera, Eric P Nawrocki, Elena Rivas, Sean R Eddy, Alex Bateman, Robert D Finn, and Anton I Petrov. 2017. "Rfam 13.0: Shifting to a Genome-Centric Resource for Non-Coding Rna Families." *Nucleic Acids Research* 46 (D1): D335–D342.

Keul, Nicholas D, Krishnadev Oruganty, Elizabeth T Schaper Bergman, Nathaniel R Beattie, Weston E McDonald, Renuka Kadirvelraj, Michael L Gross, Robert S Phillips, Stephen C Harvey, and Zachary A Wood. 2018. "The Entropic Force Generated by Intrinsically Disordered Segments Tunes Protein Function." *Nature* 563 (7732): 584.

Lorenz, Ronny, Stephan H Bernhart, Christian Hoener Zu Siederdissen, Hakim Tafer, Christoph Flamm, Peter F Stadler, and Ivo L Hofacker. 2011. "ViennaRNA Package 2.0." *Algorithms for Molecular Biology* 6 (1): 26.

Marks, Debora S, Lucy J Colwell, Robert Sheridan, Thomas A Hopf, Andrea Pagnani, Riccardo Zecchina, and Chris Sander. 2011. "Protein 3D Structure Computed from Evolutionary Sequence Variation." *PloS One* 6 (12): e28766.

Mayer, O., L. Rajkowitsch, C. Lorenz, R. Konrat, and R. Schroeder. 2007. "RNA Chaperone Activity and RNA-Binding Properties of the E. Coli Protein StpA." *Nucleic Acids Research* 35 (4): 1257–69. https://doi.org/10.1093/nar/gkl1143.

McCaskill, John S. 1990. "The Equilibrium Partition Function and Base Pair Binding Probabilities for Rna Secondary Structure." *Biopolymers: Original Research on Biomolecules* 29 (6-7): 1105–19.

Michel, Francois, and Eric Westhof. 1990. "Modelling of the Three-Dimensional Architecture of Group I Catalytic Introns Based on Comparative Sequence Analysis." *Journal of Molecular Biology* 216 (3): 585–610.

Mohr, Georg, Aixia Zhang, Janet A Gianelos, Marlene Belfort, and Alan M Lambowitz. 1992. "The Neurospora CYT-18 Protein Suppresses Defects in the Phage T4 Td Intron by Stabilizing the Catalytically Active Structure of the Intron Core." *Cell* 69 (3): 483–94.

Morcos, Faruck, Andrea Pagnani, Bryan Lunt, Arianna Bertolino, Debora S. Marks, Chris Sander, Riccardo Zecchina, José N. Onuchic, Terence Hwa, and Martin Weigt. 2011. "Direct-Coupling Analysis of Residue Coevolution Captures Native Contacts Across Many Protein Families." *Proceedings of the National Academy of Sciences* 108 (49): E1293–E1301. https://doi.org/10.1073/pnas.1111471108.

Nawrocki, Eric P., and Sean R. Eddy. 2013. "Infernal 1.1: 100-Fold Faster RNA Homology Searches." *Bioinformatics* 29 (22): 2933–5. https://doi.org/10.1093/bioinformatics/btt509.

Nawrocki, Eric P, Thomas A Jones, and Sean R Eddy. 2018. "Group I Introns Are Widespread in Archaea." *Nucleic Acids Research* 46 (15): 7970–6. https://doi.org/10.1093/nar/gky414.

Nott, Timothy J, Evangelia Petsalaki, Patrick Farber, Dylan Jervis, Eden Fussner, Anne Plochowietz, Timothy D Craggs, et al. 2015. "Phase Transition of a Disordered Nuage Protein Generates Environmentally Responsive Membraneless Organelles." *Molecular Cell* 57 (5): 936–47.

Ovchinnikov, Sergey, Hetunandan Kamisetty, and David Baker. 2014. "Robust and Accurate Prediction of Residue–Residue Interactions Across Protein Interfaces Using Evolutionary Information." *Elife* 3: e02030.

Ovchinnikov, Sergey, Lisa Kinch, Hahnbeom Park, Yuxing Liao, Jimin Pei, David E Kim, Hetunandan Kamisetty, Nick V Grishin, and David Baker. 2015. "Large-Scale Determination of Previously Unsolved Protein Structures Using Evolutionary Information." *Elife* 4: e09248.

Papaleo, Elena, Giorgio Saladino, Matteo Lambrughi, Kresten Lindorff-Larsen, Francesco Luigi Gervasio, and Ruth Nussinov. 2016. "The Role of Protein Loops and Linkers in Conformational Dynamics and Allostery." *Chemical Reviews* 116 (11): 6391–6423.

Papasaikas, Panagiotis, and Juan Valcárcel. 2016. "The Spliceosome: The Ultimate RNA Chaperone and Sculptor." *Trends in Biochemical Sciences* 41 (1): 33–45.

Paukstelis, Paul J, Jui-Hui Chen, Elaine Chase, Alan M Lambowitz, and Barbara L Golden. 2008. "Structure of a Tyrosyl-tRNA Synthetase Splicing Factor Bound to a Group I Intron RNA." *Nature* 451 (7174): 94.

Piovesan, Damiano, Francesco Tabaro, Lisanna Paladin, Marco Necci, Ivan Mičetić, Carlo Camilloni, Norman Davey, et al. 2017. "MobiDB 3.0: More Annotations for Intrinsic Disorder, Conformational Diversity and Interactions in Proteins." *Nucleic Acids Research* 46 (D1): D471–D476.

Potter, Simon C, Aurélien Luciani, Sean R Eddy, Youngmi Park, Rodrigo Lopez, and Robert D Finn. 2018. "HMMER Web Server: 2018 Update." *Nucleic Acids Research* 46 (W1): W200–W204. https://doi.org/10.1093/nar/gky448.

Reinharz, Vladimir, Yann Ponty, and Jérôme Waldispühl. 2016. "Combining Structure Probing Data on RNA Mutants with Evolutionary Information Reveals RNA-Binding Interfaces." *Nucleic Acids Research* 44 (11): e104. https://doi.org/10.1093/nar/gkw217.

Reuter, Jessica S, and David H Mathews. 2010. "RNAstructure: Software for RNA Secondary Structure Prediction and Analysis." *BMC Bioinformatics* 11 (1): 129.

Sayers, Eric W, Mark Cavanaugh, Karen Clark, James Ostell, Kim D Pruitt, and Ilene Karsch-Mizrachi. 2019. "GenBank." *Nucleic Acids Research* 47 (D1): D94–D99. https://doi.org/10.1093/nar/gky989.

Schad, Eva, Erzsébet Fichó, Rita Pancsa, István Simon, Zsuzsanna Dosztányi, and Bálint Mészáros. 2017. "DIBS: A Repository of Disordered Binding Sites Mediating Interactions with Ordered Proteins." *Bioinformatics* 34 (3): 535–37.

Schuwirth, Barbara S, Maria A Borovinskaya, Cathy W Hau, Wen Zhang, Antón Vila-Sanjurjo, James M Holton, and Jamie H Doudna Cate. 2005. "Structures of the Bacterial Ribosome at 3.5 Å Resolution." *Science* 310 (5749): 827–34.

Tompa, Peter, and Peter Csermely. 2004. "The Role of Structural Disorder in the Function of RNA and Protein Chaperones." *The FASEB Journal* 18 (11): 1169–75.

Toth-Petroczy, Agnes, Perry Palmedo, John Ingraham, Thomas A Hopf, Bonnie Berger, Chris Sander, and Debora S Marks. 2016. "Structured States of Disordered Proteins from Genomic Sequences." *Cell* 167 (1): 158–70.

Van Der Lee, Robin, Marija Buljan, Benjamin Lang, Robert J Weatheritt, Gary W Daughdrill, A Keith Dunker, Monika Fuxreiter, et al. 2014. "Classification of Intrinsically Disordered Regions and Proteins." *Chemical Reviews* 114 (13): 6589–6631.

Varadi, Mihaly, and Peter Tompa. 2015. "The Protein Ensemble Database." In *Intrinsically Disordered Proteins Studied by Nmr Spectroscopy*, 335–49. Springer.

Waldsich, Christina, Rupert Grossberger, and Renée Schroeder. 2002. "RNA Chaperone StpA Loosens Interactions of the Tertiary Structure in the Td Group I Intron in Vivo." *Genes & Development* 16 (17): 2300–2312. https://doi.org/10.1101/gad.231302.

Weigt, Martin, Robert A. White, Hendrik Szurmant, James A. Hoch, and Terence Hwa. 2009. "Identification of Direct Residue Contacts in Proteinprotein Interaction by Message

Passing." *Proceedings of the National Academy of Sciences* 106 (1): 67–72. https://doi.org/10.1073/pnas.0805923106.

Weinreb, Caleb, Adam J Riesselman, John B Ingraham, Torsten Gross, Chris Sander, and Debora S Marks. 2016. "3D RNA and Functional Interactions from Evolutionary Couplings." *Cell* 165 (4): 963–75.

Woodson, Sarah A. 2010. "Taming Free Energy Landscapes with RNA Chaperones." *RNA Biology* 7 (6): 677–86.

Wright, Peter E, and H Jane Dyson. 2015. "Intrinsically Disordered Proteins in Cellular Signalling and Regulation." *Nature Reviews Molecular Cell Biology* 16 (1): 18.

Zhou, Yu, Chen Lu, Qi-Jia Wu, Yu Wang, Zhi-Tao Sun, Jia-Cong Deng, and Yi Zhang. 2008. "GISSD: Group I Intron Sequence and Structure Database." *Nucleic Acids Research* 36 (suppl_1): D31–D37. https://doi.org/10.1093/nar/gkm766.

a.

b.

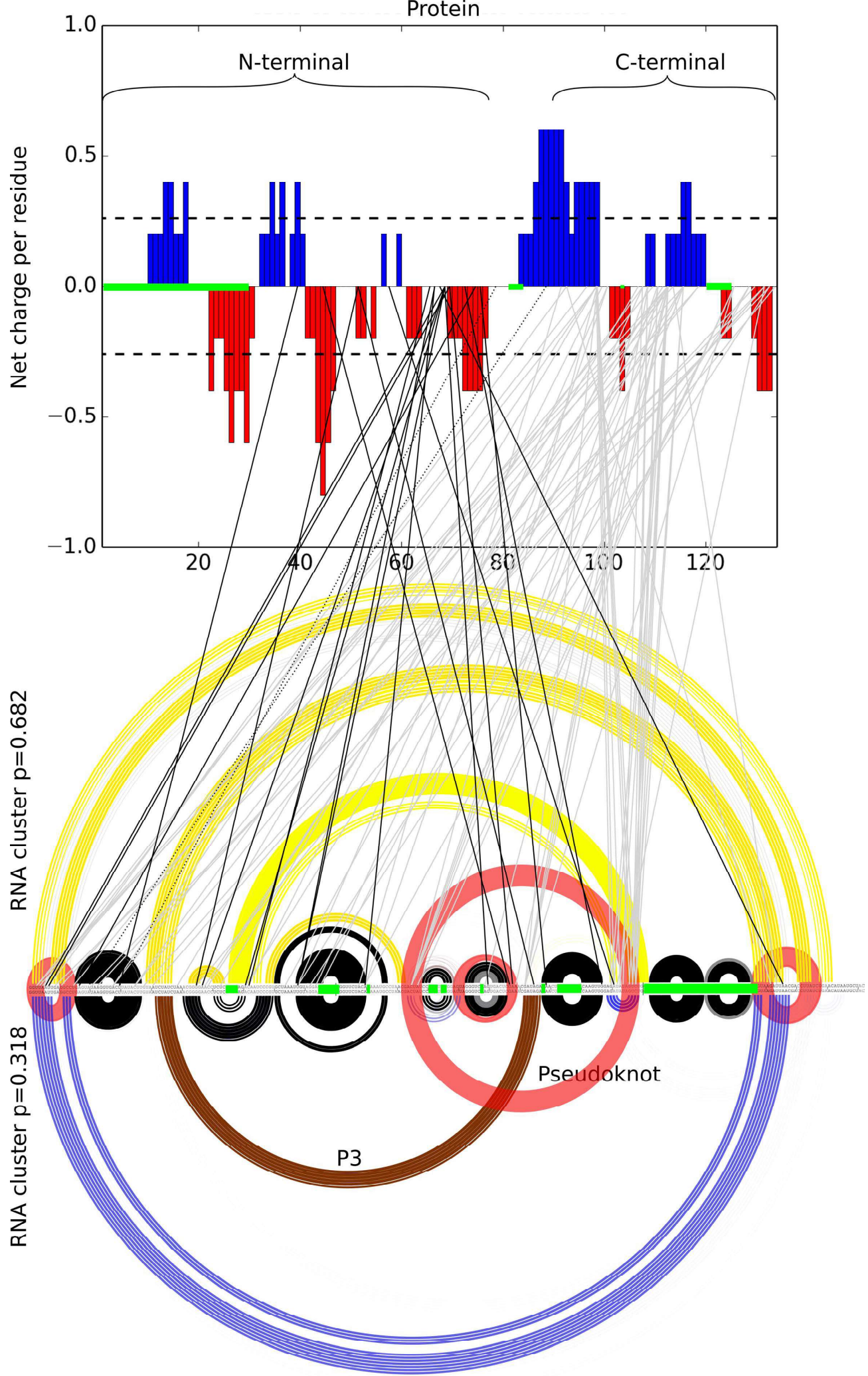

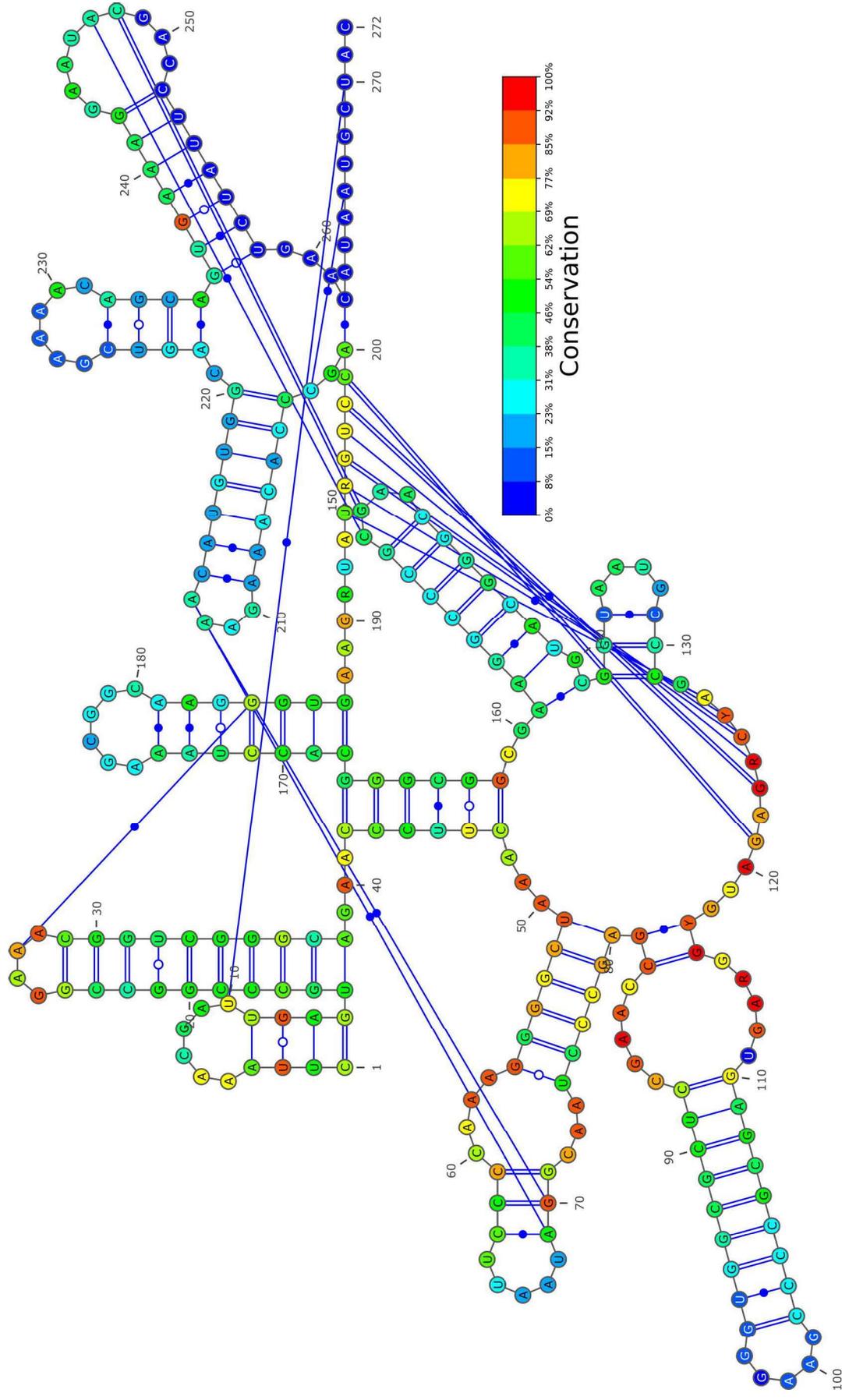

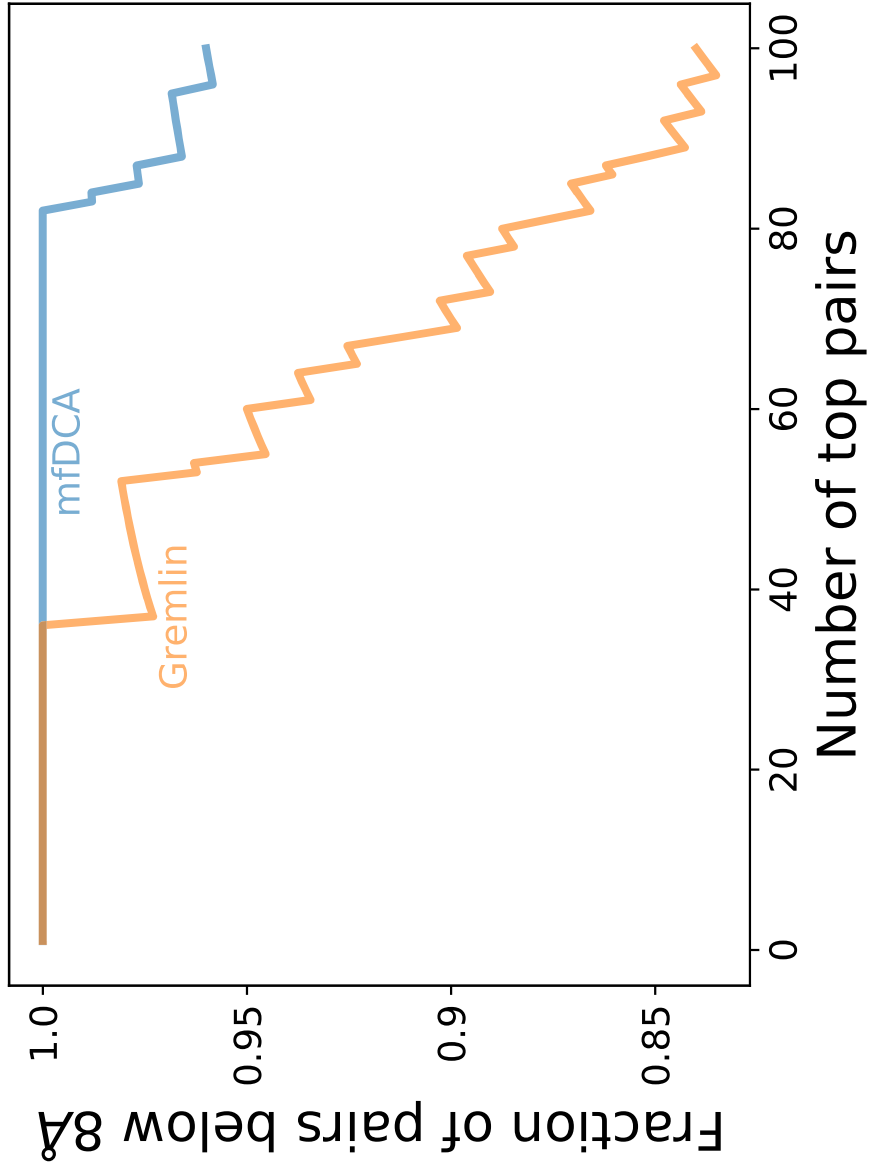

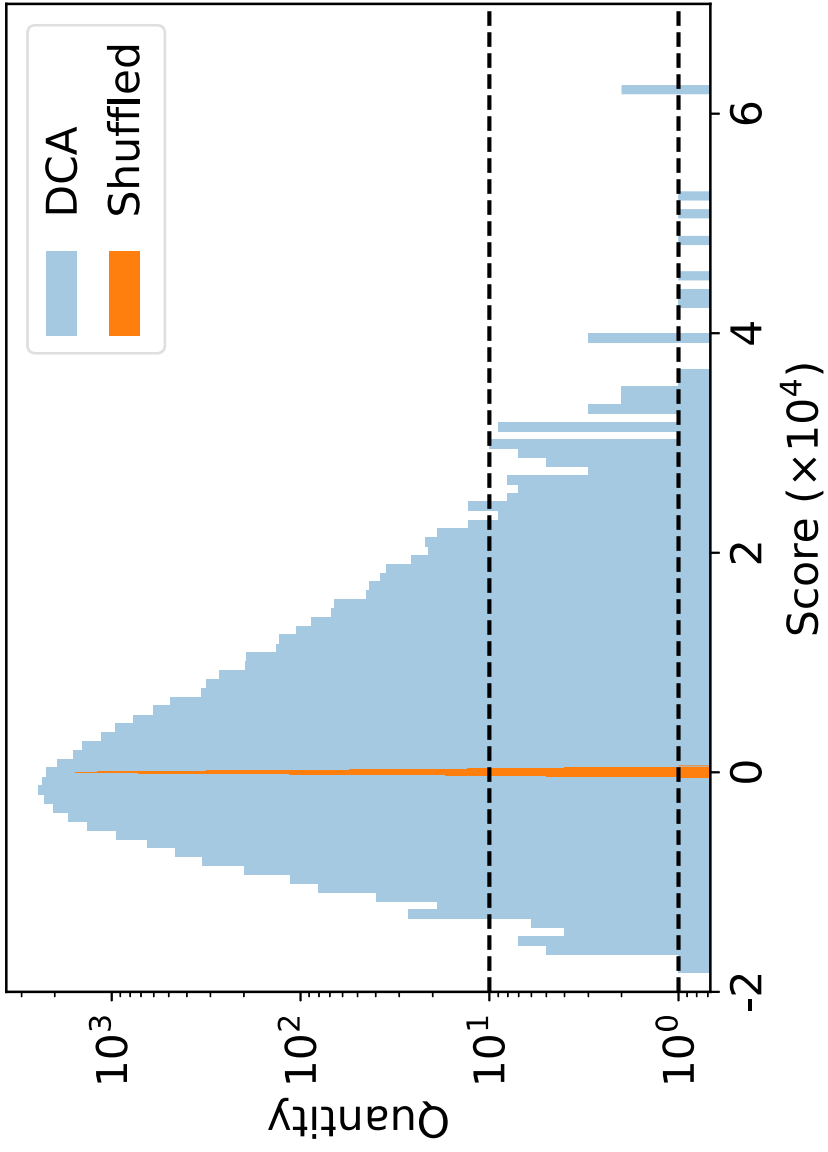

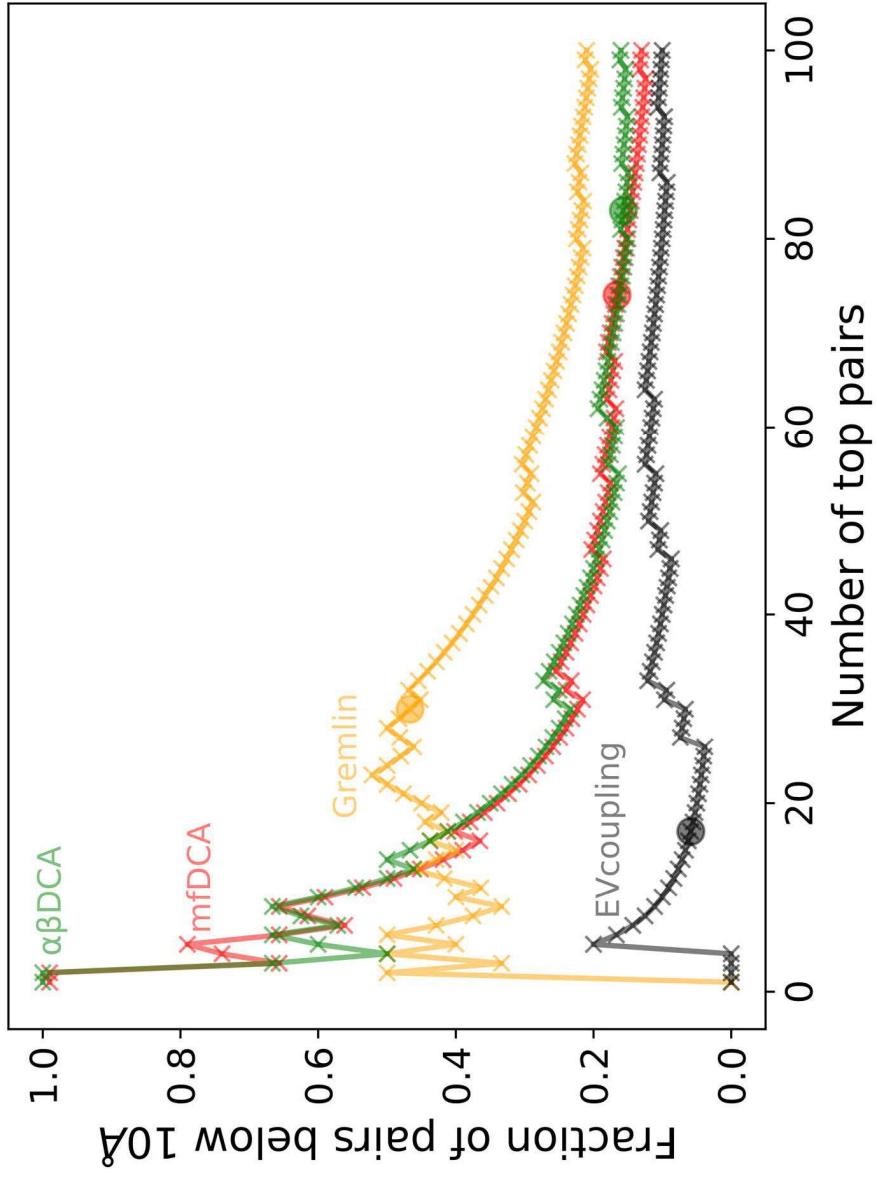

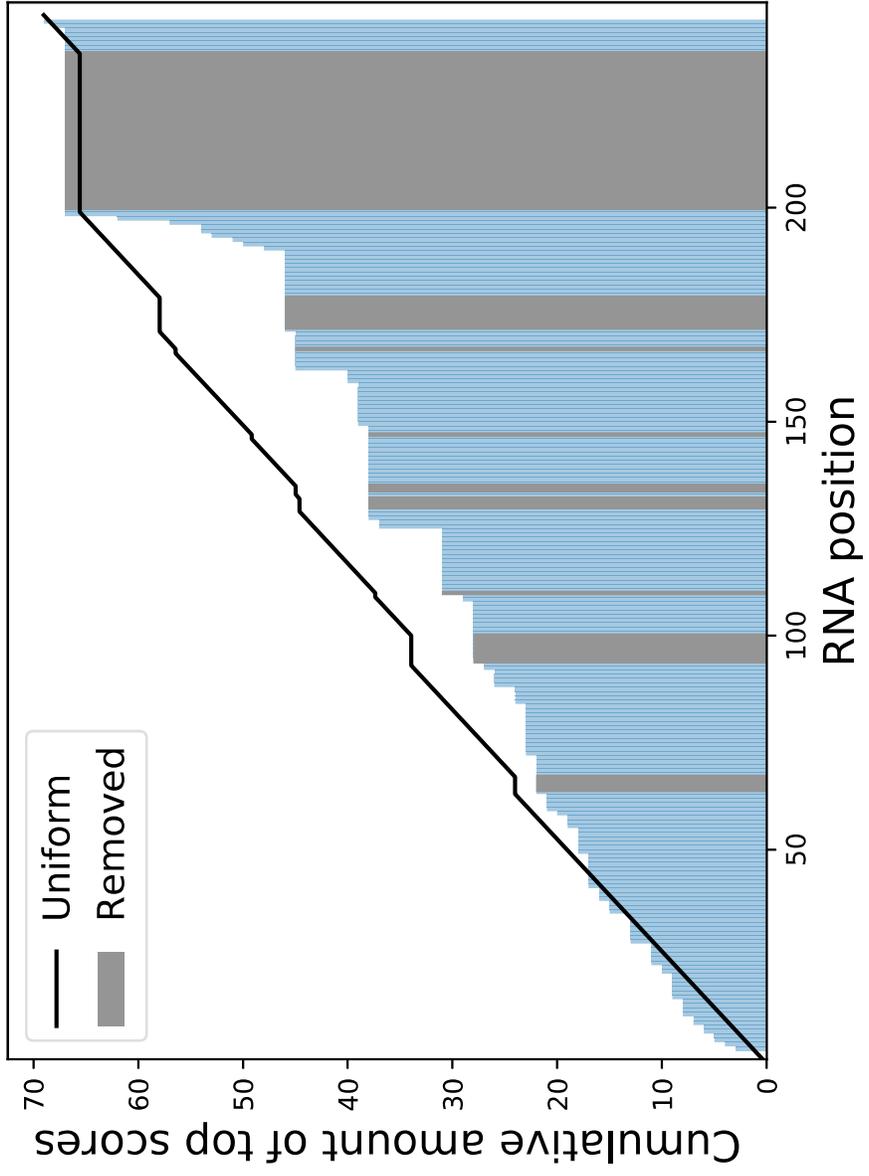

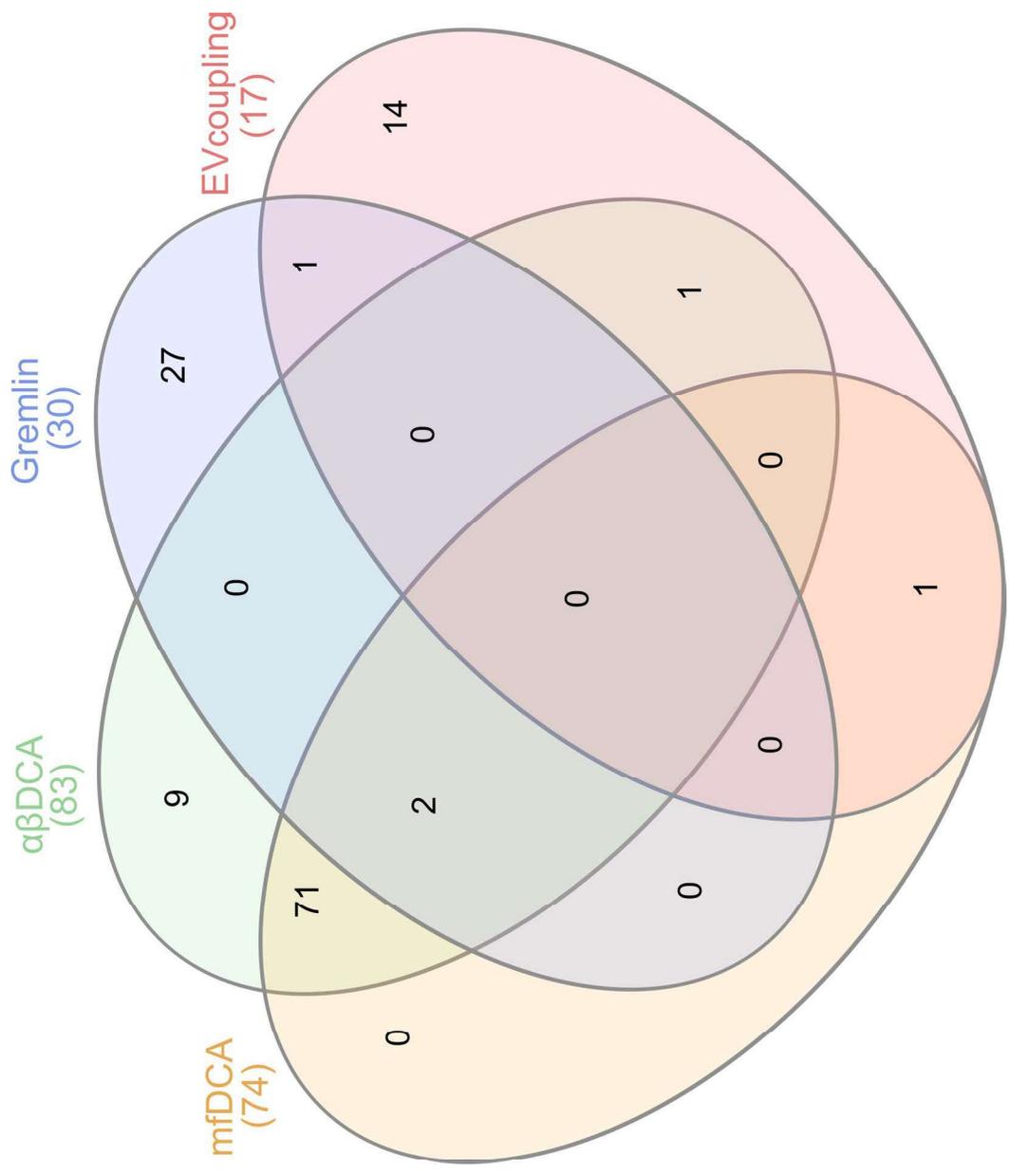